# From layered 2D carbon to 3D tetrahedral original allotropes $C_{12}$ and $C_{18}$ with physical properties related to diamond: Crystal chemistry and DFT investigations.

Samir F. Matar, Lebanese German University, Sahel-Alma, Jounieh. Lebanon

ORCID: https://orcid.org/my-orcid?orcid=0000-0001-5419-358X

Email: s.matar@lgu.edu.lb

*Abstract*

*Two mechanisms of changes from 2D to 3D (D = dimensionality) involving 2D $C(sp^2)$ trigonal paving to $C(sp^3)$ tetrahedral stacking are proposed through puckering of the 2D layers on one hand and interlayer insertion of extra C on the other hand. Such transformations, led to original 3D hexagonal $C_{12}$ and $C_{18}$ allotropes respectively characterized by **lon** and **bac** topologies. Using density functional theory DFT calculations, the two allotropes were found cohesive and stable both mechanically (elastic properties) and dynamically (phonons). Comparisons of the physical properties with known **uni** $C_6$ were established letting identify ranges of large Vickers hardness: $H_V$ (**uni** $C_6$) = 89 GPa, $H_V$ (**lon** $C_{12}$) = 97 GPa, and $H_V$ (**bac** $C_{18}$) = 70 GPa. Whilst $C_6$ was identified with acoustic phonons instability, $C_{12}$ and $C_{18}$ were found stable dynamically throughout the acoustic and optic frequency ranges. Furthering on the thermal properties the allotropes were characterized with a temperature dependence curve of the specific heat $C_V$ close to experimental data of diamond with best fit for novel $C_{18}$. The electronic band structures reveal a small band gap of 1 eV for **uni** $C_6$ and larger direct band gap of 3 eV for the two other 3D allotropes. Such modulations of the electronic and physical properties should open scopes of carbon research.*

Keywords: Carbon allotropes; DFT; hardness; phonons; electronic structure

## 1- Introduction.

The change from 2D graphitic-like carbon to 3D diamond-like structure involves physical and chemical mechanisms that have been studied both experimentally and through different levels of modeling over the last 25 years (cf. [1] and therein cited works). In 1999 Fayos showed that successive contractions of graphite in different directions, accompanied with $C(sp^2)$ to $C(sp^3)$ hybridization changes stimulate a graphite to diamond (G2D) phase transition [2]. Also, a synergy between experiment and simulation at the atomistic level and under high pressure/high temperature, let investigate the G2D mechanism with the observation of cubic diamond (CD) involving a preferred growth of hexagonal diamond (HD) appearing only as twin structures of CD [3]. Later, Guo et al. studied such mechanism in nano-crystalline graphite with shear-promoted G2D change occurring at the grain boundary [4]. Highlighting topology aspects, Öhrström and O'Keeffe [5] proposed a relevant approach to new allotropes of the group 14 elements with labels such as **dia**, for CD, **lon**, for HD, etc…. The changes of geometries and atomic re-arrangements involve topology are analyzed thanks to TopCryst program [6]. Particularly for Carbon allotropes the ongoing interest within the scientific community involved with experimental and theoretical research, the well-known and the novel allotropes are listed in the SACADA database available online [7].

In this context, the present investigation proposes to broaden the G2D-like changes. Going from 2D $C(sp^2)$ paving to 3D $C(sp^3)$ tetrahedral arrangements is proposed to be initiated by:

- the puckering of atomic layers through driving carbon into off-plane position, on one hand, and
- a rationalized interlayer insertion of extra C to enforce the 3D structure, on the other hand.

The crystal structure manipulations are systematically followed by geometry optimizations quantum mechanics calculations onto ground state structures followed by energy derived physical properties particularly those reporting on the mechanical and dynamic ones.

## 2- Computational methodology

The determination of the ground state structures corresponding to the energy minima and the prediction of their mechanical and dynamical properties were carried out within the widely accepted framework of the density functional theory DFT. The DFT was initially proposed in two iconic publications: in 1964, Hohenberg and Kohn developed the theoretical framework [8], and in 1965, Kohn and Sham established the Kohn-Sham equations for practical solution of the wave equation [9].

Based on the DFT, calculations were performed within the Vienna Ab initio Simulation Package (VASP) code [10,11] and the Projector Augmented Wave (PAW) method [11,12] for the atomic potentials. DFT exchange correlation (XC) effects were considered using the generalized gradient approximation (GGA) [13]. Preliminary calculations with the native DFT-XC local density approximation (LDA) [14] resulted in underestimated lattice constants at ambient pressure and were therefore abandoned. Relaxation of the atoms to the ground state structures was performed with the conjugate gradient algorithm according to Press *et al.* [15]. The Bloechl tetrahedron method [16] with corrections according to the Methfessel and Paxton scheme [17] was used for geometry optimization and energy calculations, respectively. Brillouin-zone (BZ) integrals were approximated by a special **k**-point sampling according to Monkhorst and Pack [18]. Structural parameters were optimized until atomic forces were below 0.02 eV/Å and all stress components were < 0.003 eV/Å$^3$. The calculations were converged at an energy cutoff of 400 eV for the plane-wave basis set in terms of the **k**-point integration in the reciprocal space from $k_x(6) \times k_y(6) \times k_z(4)$ up to $k_x(12) \times k_y(12) \times k_z(8)$ to obtain a final convergence and relaxation to zero strains for the original stoichiometries presented in this work. In the post-processing of the ground state electronic structures, the charge density projections were operated on the lattice sites.

The mechanical stabilities were obtained from the elastic constants' calculations. The treatment of the results was done using ELATE online tool devoted to the analysis of the elastic tensors [19]. The program provides the bulk (*B*) and shear (*G*) modules along different averaging methods; the Voigt method [20] was used here. Two methods of microscopic theory of hardness by Tian et al. [21] and Chen et al. [22] were used to estimate the Vickers hardness ($H_V$) from the elastic constants.

The dynamic stabilities were confirmed by the positive phonon magnitudes. The corresponding phonon band structures were obtained from a high resolution of the hexagonal

Brillouin zone according to Togo *et al.* [23]. Experimental specific heat $C_V$ data of diamond needed to assess the calculated results of the three 3D allotropes were obtained from Victor works [24]. The energy volume equations of states EOS used to assess the 2D/3D $C_6$ allotropes were obtained with the fit of the E(V) curves with Birch EOS [25]. The electronic band structures were obtained using the all-electron DFT-based ASW method [26] and the GGA XC functional [13]. The VESTA (Visualization for Electronic and Structural Analysis) program [27] was used to visualize the crystal structures and charge densities.

### 3- Crystal chemistry and energy-volume equations of state

*3-1 Hexagonal 2D/3D forms of $C_6$.*

Among the highlighted topology labels listed by Öhrström and O'Keeffe [5] such as **dia**, for CD, **lon**, for HD, the **uni** topology points to hexagonal 3D **$C_6$** (SG $P6_122$, No. 178) with corner sharing tetrahedra as shown in Fig. 1a. **uni** $C_6$ is listed under No. 56 in SACADA. Interestingly, the 3D structure is preserved with the higher symmetry SG, $P6_522$, No. 179. The calculation of the ground state lattice parameters and atomic positions shown in the 1$^{st}$ column of Table 1 are found close to the structure in SACADA database. The density is large: $\rho = 3.44$ g/cm$^3$, approaching diamond's with $\rho = 3.54$ g/cm$^3$, letting expect a high hardness. Furthering on crystal symmetry analysis, we considered other space groups: namely *P*622 No. 177 and *P*6*mm* No. 183 resulting both in a 2D $C_6$. The structure obtained with the latter SG was subsequently submitted to calculations of the ground state configuration. The 2$^{nd}$ column showing the 2D-$C_6$ data reveal larger crystal parameters and volume with $x_C = 2/3$, i.e., relevant to perfect hexagons forming the layer (Fig. 1b), whereas $x_C$ amounted to a lower value of 0.617 –off 2/3– in pristine 3D $C_6$. The topology is now **hcb**, not documented as such in SACADA database but a correspondence with **sqc**1427 (SACADA No. 125) was found for 2D $C_{16}$ in orthorhombic symmetry, citing Fayos' work [2]. The density is much lowered with $\rho = 2.22$ g/cm$^3$ because of the increased volume leading to a more cohesive allotrope with $E_{coh.}$/atom= -2.59 eV versus -1.76 for **uni** $C_6$. These results and observations let propose a room pressure, large volume, low density **hcb** 2D-$C_6$ *versus* a high pressure, smaller volume, higher density **uni** $C_6$. These statements were verified with establishing the corresponding energy-volume equations of state (EOS) and the fit parameters to evaluate the equilibrium values. For the purpose a series of calculations of the total energy as a function of volume were carried out for 2D and 3D $C_6$ The resulting $E(V)$ curves, shown in Figure 2. The curves were fitted to the 3$^{rd}$ order Birch equation of state [25]:

$$E(V) = E_0(V_0) + (9/8) \cdot V_0 B_0 [([(V_0)/V])^{2/3} - 1]^2 + (9/16) \cdot B_0 \cdot (B' - 4) \cdot V_0 [([(V_0)/V])^{2/3} - 1]^3,$$

where $E_0$, $V_0$, $B_0$, and $B'$ are the equilibrium energy; volume; bulk modulus; and its first pressure derivative, respectively.

The fit results in the insert illustrate the observations and let expect a possible G2D like transition involving a decrease of cell volume and increase of energy. The volume at the crossing of the two curves occurs at 5 Å$^3$ below the equilibrium volume of 35 Å$^3$ signaling a potential phase transition under reasonable pressures.

### *3-2 Devising carbon rich 2D template structures*

Nevertheless, such devised 2D $C_6$ structure could not be used to G2D-like transformation through puckering (out-of-plane C) neither through carbon insertion as planned for this work. Further crystallographic investigations let resolve the structure with the same coordinates ($z_C$ = 2/3) into **hcb**-$C_{12}$ in higher symmetry SG $P6_3/mcm$ No. 193 with a unique carbon site C (12k) 2/3, 0, ½. The ground state multilayer structure shows a twice larger volume but a significant densification due to the larger number of stacking layers as shown in Fig. 1c. The puckering of $C_{12}$ layers leads for a threshold value of $z_C$ = 0.0628 to 3D **lon** $C_{12}$ that was subsequently geometry optimized. The 4$^{th}$ column of results in Table 1 shows an increase of the cohesive energy with respect to 2D $C_{12}$ with -2.47 eV/at. close to the value obtained for diamond: -2.49 eV/at. The density is also close to diamond, and it is expected that **lon** $C_{12}$ should behave alike diamond for the physical properties. The structure shown in Fig. 1d exhibits regular corner sharing tetrahedra as well as tetrahedra in two layers connecting via segments as illustrated with the polyhedral projection.

Lastly, along with the 2$^{nd}$ working hypothesis presented in the introduction, starting from 2D $C_{12}$, the insertion of carbon at (6g) position of the same space group No. 193, i.e., with x, 0, ¼, led to $C_{18}$ with smaller cohesive energy as well as density *versus* **lon** $C_{12}$. $C_{18}$ was identified with **bac** topology using TopCryst. Such topology is undocumented in SACADA database. From the atomic positions (Table 1 last column), it can be observed that the insertion of carbon involves almost twice smaller $z_C$ coordinate, letting suggest that the insertion compensates for puckering of 2D $C_{12}$ needed for the formation of **lon** $C_{12}$. The structure is shown in Fig. 1e where the inserted carbon at (*6g*) positions is shown with white spheres. Such 3D structure exhibits corner sharing tetrahedra as well as tetrahedra connecting via interlayers C-C segments as in **lon** $C_{12}$.

4- **Mechanical properties from the elastic properties**

The investigation of the mechanical properties was based on the calculations of the elastic properties determined by performing finite distortions of the lattice and deriving the elastic constants from the strain-stress relationship. The calculated sets of elastic constants $C_{ij}$ (i and j indicate directions) are given in Table 2a. All $C_{ij}$ values are positive signaling stability of the three allotropes. Using ELATE program introduced above [19], the bulk (*B*) and the shear (*G*) modules obtained by averaging the elastic constants using Voigt's method are given in Table 2b.

The largest values of bulk and shear modules are found for **lon** $C_{12}$ with $B_V$ =444 GPa and $G_V$= 534 GPa, close to the accepted values for diamond, $B_V$ =444 GPa and $G_V$ = 550 GPa [28]. Lower values are observed for **uni**-$C_6$ and then **bac**-$C_{18}$. The Vickers hardness $H_V$ was then calculated with two models of microscopic theory of hardness:

$H_V= 0.92(G_V/B_V)^{1.137} G_V^{0.708}$ (Tian et al.) [21]

$H_V= 2(G^3/B^2)^{0.585}-3$ (Chen et al.) [22]

The corresponding Vickers hardness ($H_V$) magnitudes obtained along the two methods are given in the last two columns of Table 2b. They show close magnitudes. The largest hardness is obtained for **lon**-$C_{12}$ with magnitudes close to diamond. Then **uni**-$C_6$ with close values of hardness along the two approaches can be assigned ultra-hard behavior, alike diamond. Oppositely, **bac**-$C_{18}$ has the lowest magnitude of $H_V$=70 GPa letting consider it with as super hard allotrope.

5- **Dynamic and thermodynamic properties**

*5.1 Phonons band structures*

To verify the dynamic stability of the 3D carbon allotropes, an analysis of their phonon properties was performed. The phonon band structures obtained from a high resolution of the hexagonal Brillouin zone BZ in accordance with the method proposed by Togo *et al*. [23] are shown in Fig. 2. The BZ is depicted in Fig. 2a [29]. In the three panels (b-d) the bands (red lines) develop along the main directions of hexagonal Brillouin zone (horizontal *x*-axis), separated by vertical lines for better visualization, while the vertical direction (*y*-axis) represents the frequencies ω, given in terahertz (THz).

The band structures include 3N bands describing three acoustic modes starting from zero energy (ω = 0) at the Γ point (the center of the Brillouin zone) and reaching up to a few terahertz, and 3N-3 optical modes at higher energies. The low-frequency acoustic modes are associated with the rigid translation modes (two transverse and one longitudinal) of the crystal lattice. The calculated phonon frequencies are all positive, indicating that the three 3D allotropes are dynamically stable. Nevertheless, an anomaly is shown in Fig. 2b relative to **uni** $C_6$: whilst along Γ−K direction featuring in-plane vibrations positive frequencies are observed, slight negative frequencies appear along the direction Γ−A relevant to vibrations along vertical direction (Fig. 2a). This is not observed for the two newly found $C_{12}$ and $C_{18}$, letting deduct some dynamic instability for **uni** $C_6$, oppositely to the two other allotropes found dynamically stable. In all three carbon allotropes, besides the dispersed bands, the highest bands are observed around ~40 THz close to the value observed for diamond by Raman spectroscopy [30], letting suggest a relationship with diamond. Such an assumption needs to be supported by further analysis of the thermal properties of the three allotropes in comparison to diamond.

### *5.2 Temperature dependence of the heat capacity*

The thermodynamic properties of the three allotropes were calculated from the phonon frequencies using the statistical thermodynamic approach [31] on a high-precision sampling mesh in the hexagonal BZ. The temperature dependencies of the heat capacity at constant volume ($C_V$) are presented in Figure 3 in comparison with the available experimental $C_V$ data for diamond by Victor [32]. Whilst all three calculated curves show good shape agreement with diamond's experimental points (red circles), the best fit is found for **bac** $C_{18}$ followed by **lon** $C_{12}$ with a higher curved and lastly **uni** $C_6$. It can be concluded that all three allotropes have close thermal properties with diamond.

### 6- **Electronic band structures**

Using the crystal parameters in Table 1, the electronic band structures were obtained for the 3D carbon allotropes using the all-electrons DFT-based augmented spherical wave method (ASW) [23] and GGA XC approximation [13]. The band structures are displayed in Figure 4.

The bands develop along the main directions of the primitive hexagonal Brillouin zone (Fig. 2a). Along the vertical direction all three panels exhibit an energy gap signaling a semi-conducting behavior. The zero energy is then considered at the top of the valence band, $E_V$. In Fig. 4a representing $C_6$ the band gap is close to 1 eV, whereas it is larger, with 3 eV magnitude for the other two allotropes $C_{12}$ and $C_{18}$ occurring between $\Gamma_V$ and $\Gamma_C$. Note that the 3 eV gap magnitude is 2 eV short of the band gap of diamond letting suggest interesting semi-conductor properties, should they be potentially prepared.

**Conclusion**

The purpose of this investigation was to present 'graphite-to-diamond' G2D-like changes of carbon crystal systems from 2D $C(sp^2)$ plane paving to 3D $C(sp^3)$ tetrahedral. The schematics goes along two mechanisms pertaining to inducing puckering of the 2D layers, on one hand, and through interlayer insertions of extra C to enforce the 3D structure, on the other hand. 2D and 3D hexagonal allotropes were used to propose original results whereby original 3D $C_{12}$ and $C_{18}$ with **lon** and **bac** topologies were characterized with mechanical, dynamic, and thermodynamic properties closely related to diamond.

**Declaration:**

I, author, declare that there are no known conflicts of interest associated with this publication and there has been no significant financial support for this work that could have influenced its outcome.

I confirm that the manuscript has been read carefully, no co-authors.

I confirm that I have given due consideration to the protection of intellectual property associated with this work and that there are no impediments to publication, including the timing of publication, with respect to intellectual property. In so doing I confirm that I have followed the regulations of my institution concerning intellectual property.

TABLES

Table 1  Crystal structure transformations of hexagonal carbon from 2D to 3D. For uni-C6 present work calculated parameters are between brackets

|  | uni-$C_6$<br>$P6_122$ (No. 178)<br>**3D** | hcb-$C_6$<br>$P6/mmm$ (No. 191)<br>**2D** | hcb-$C_{12}$<br>$P6_3/mcm$ (No. 193)<br>**2D** | **lon** $C_{12}$<br>$P6_3/mcm$ (No. 193)<br>3D | **bac** $C_{18}$<br>$P6_3/mcm$ (No. 193)<br>3D |
|---|---|---|---|---|---|
| $a$, Å | 3.996 (4.038) | 4.2654 | 4.2619 | 4.3390 | 4.2891 |
| $c$, Å | 2.444 (2.454) | 3.4289 | 6.6800 | 5.581 | 7.0029 |
| Volume, Å$^3$ | 33.79 (34.66) | 54.0258 | 105.07 | 67.98 | 111.57 |
| Density $\rho$ (g/cm$^3$) | 3.44 | 2.22 | 2.47 | 3.51 | 3.22 |
| Atomic positions | C (6a)<br>0.617 (0.616), 0, 0 | C (6d)<br>2/3, 0, ½ | C (12k)<br>2/3, 0, ½ | C (12k)<br>2/3, 0, 0.0628 | C1 (12k)<br>0.6709, 0, 0.0393<br>C2 (6g) 0.7903, 0, ¼ |
| $E_{total}$, eV<br>$E_{coh}$/atom, eV | -50.54<br>-1.76 | −55.18<br>−2.59 | -110.18<br>-2.58 | −108.83<br>−2.47 | -155.59<br>-2.04 |

N.B. E(C) = -6.6 eV.  $E_{coh}$/atom (diamond) = -2.49 eV.

Table 2  Elastic constants ($C_{ij}$) of the 3D carbon allotropes and the Hardness using Voigt averaging methods using tow semi-empirical models (all values are in GPa).

a) Elastic constants

|  | $C_{11}$ | $C_{12}$ | $C_{13}$ | $C_{33}$ | $C_{44}$ | $C_{66}$ |
|---|---|---|---|---|---|---|
| $C_6$ | 986 | 120 | 24 | 1307 | 433 | 467 |
| $C_{12}$ | 1204 | 106 | 14 | 1324 | 549 | 462 |
| $C_{18}$ | 871 | 78 | 12 | 1216 | 397 | 293 |

b) Bulk and shear moduli in GPa units and resulting hardness within two approaches (cf. Text).

|  | $B_V$ | $G_V$ | $H_{V\,Tian}$ | $H_{V\,Chen}$ |
|---|---|---|---|---|
| $C_6$ | 402 | 481 | 89 | 89 |
| $C_{12}$ | 444 | 534 | 97 | 95 |
| $C_{18}$ | 351 | 387 | 70 | 70 |

FIGURES

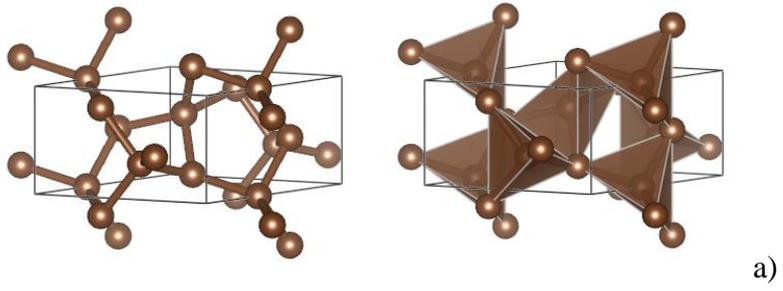

a)

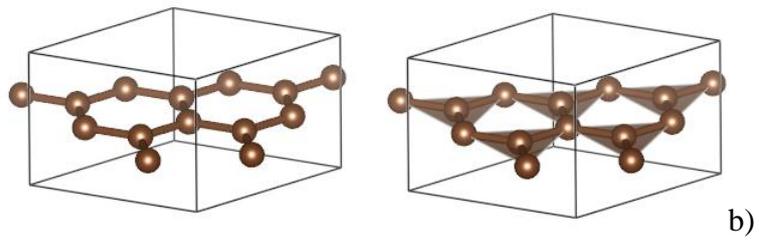

b)

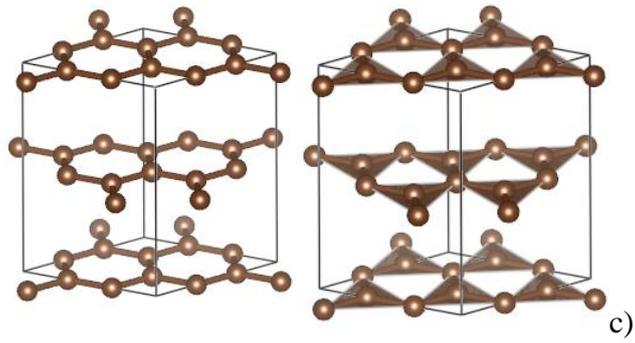

c)

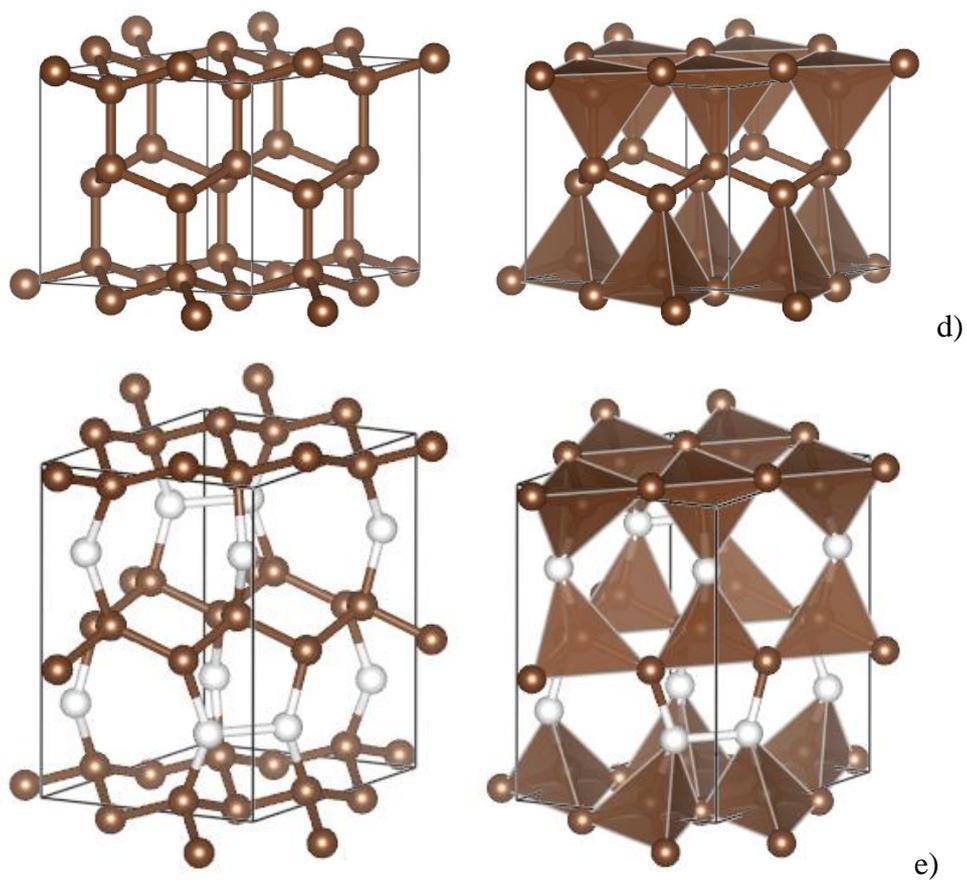

Figure 1. Sketches of the carbon allotropes in ball and stick and polyhedral representations: a) 3D **uni** $C_6$, b) 2D **hcb** $C_6$, c) 2D **hcb** $C_{12}$, d) 3D **lon** $C_{12}$, e) 3D **bac** $C_{18}$ (the white speheres represent the inserted carbon in 2D-$C_{12}$, cf. text).

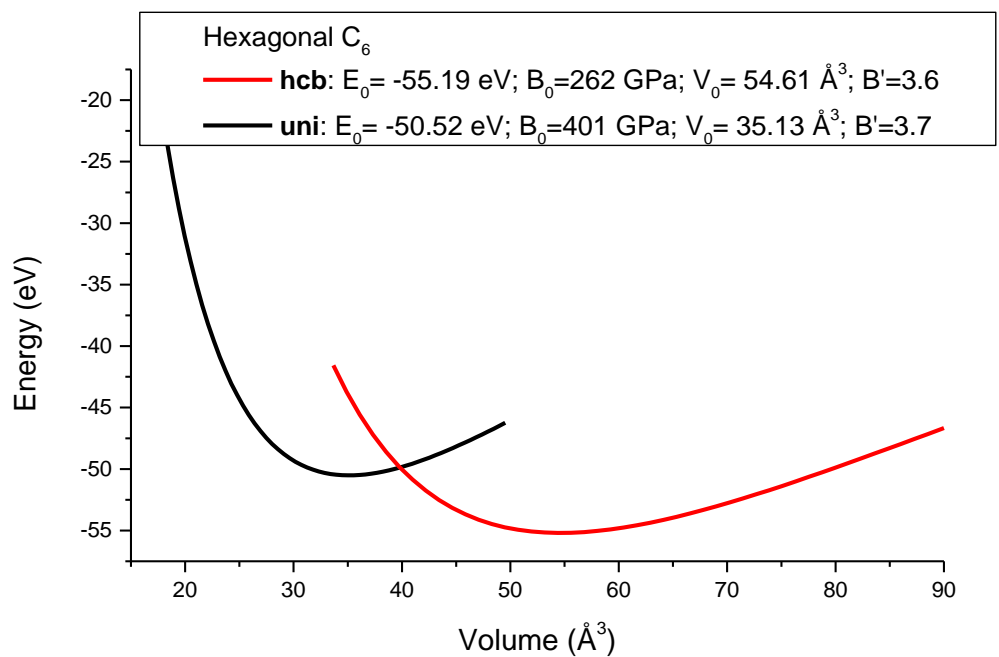

Figure 2. Hexagonal $C_6$ in 2D (**hcb**) and 3D (**uni**) forms: Energy-volume curves and fit values using Birch equation of state.

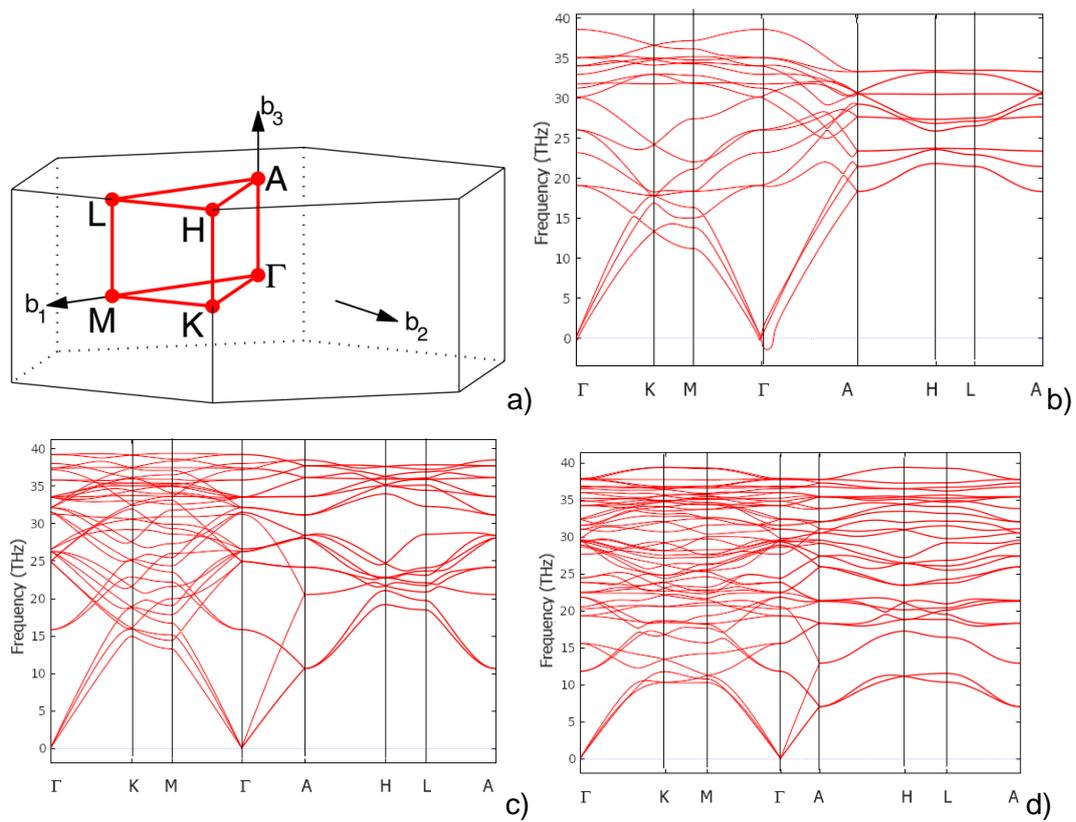

Figure 3. Phonons band structures of the carbon allotropes along major lines of the hexagonal Brillouin zone (a) of **uni** $C_6$ (b), **lon** $C_{12}$ (c), and **bac** $C_{18}$ (d).

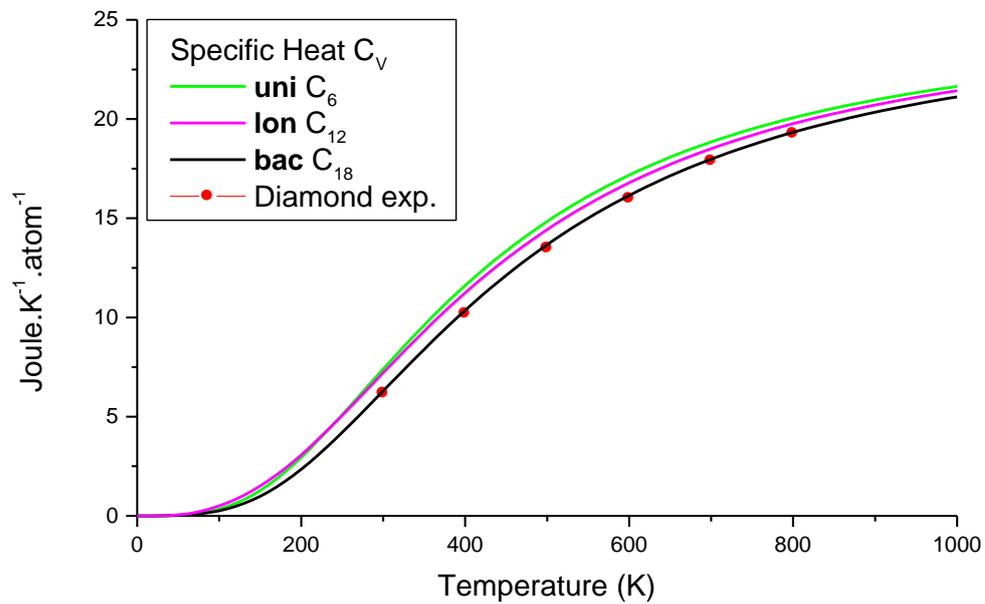

Figure 4. Temperature dependence of the specific heat $C_V$.

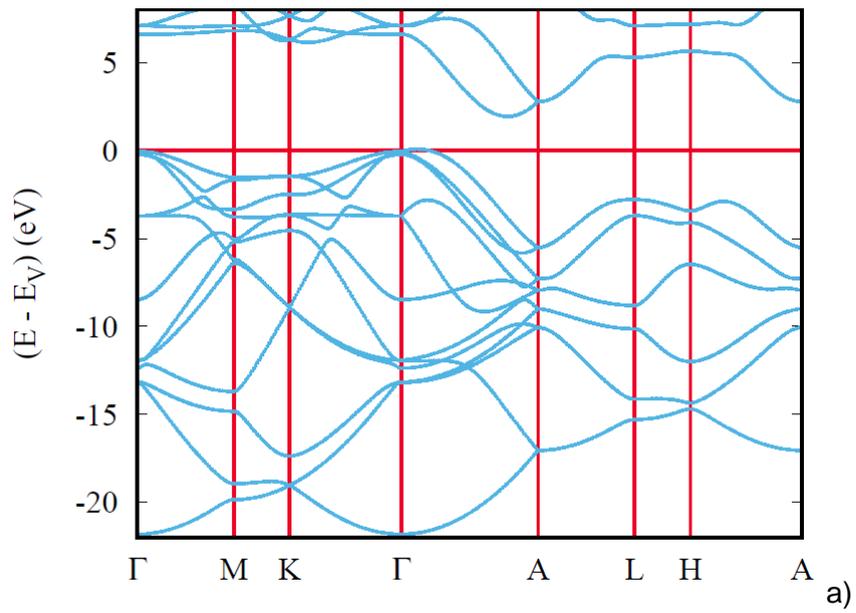

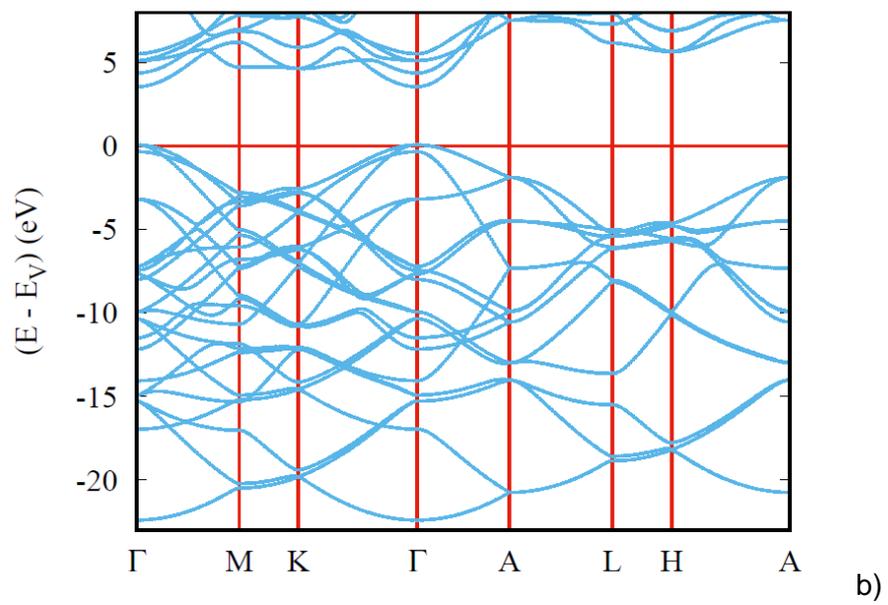

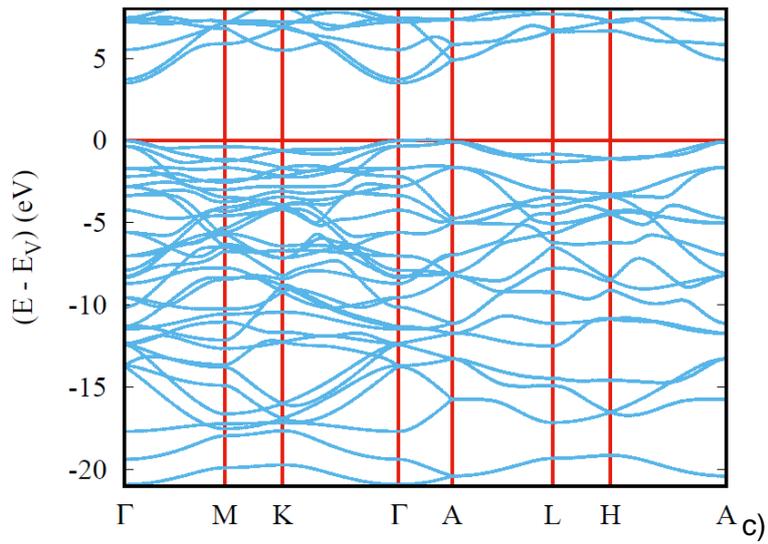

Figure 5. Electronic band structures of the 3D carbon allotropes along major lines of the hexagonal Brillouin zone. a) 3D **uni** $C_6$, b) 3D **lon** $C_{12}$, c) 3D **bac** $C_{18}$.